\documentclass{aa} 
\usepackage{graphicx, amssymb,epsfig} 
\usepackage{mathptm}
\usepackage{times}
\usepackage{txfonts}
\usepackage{multirow}
\usepackage{natbib}
\usepackage{longtable}
\usepackage{color}
\usepackage[outercaption]{sidecap}
\usepackage[usenames,dvipsnames]{xcolor}
\begin{document} 
\title{A hidden reservoir of Fe/FeS in interstellar silicates?}
\author{M. K{\"o}hler\and
             A. Jones \and
             N. Ysard}  

  \institute{Institut d'Astrophysique Spatiale (IAS), Universit\'e Paris Sud \& CNRS, B\^at. 121, 
       Orsay 91405, France  } 
 
\date{Received ..; accepted ..} 

\abstract
  {The depletion of iron and sulphur into dust in the interstellar medium and the exact nature of interstellar amorphous silicate grains is still an open question.}
  {We study the incorporation of iron and sulphur into amorphous silicates of olivine- and pyroxene-type and their effects on the dust spectroscopy and thermal emission.}
   {We used the Maxwell-Garnett effective-medium theory to construct the optical constants for a mixture of silicates, metallic iron, and iron sulphide. We also studied the effects of iron and iron sulphide in aggregate grains.}
   {Iron sulphide inclusions within amorphous silicates that contain iron metal inclusions shows no strong differences in the optical properties of the grains. A mix of amorphous olivine- and pyroxene-type silicate broadens the silicate features. An amorphous carbon mantle with a thickness of 10 nm on the silicate grains leads to an increase in absorption on the short-wavelength side of the 10\,$\mu$m silicate band.}
   {The assumption of amorphous olivine-type and pyroxene-type silicates and a 10 nm thick amorphous carbon mantle better matches the interstellar silicate band profiles. 
  Including iron nano-particles leads to an increase in the mid-IR extinction, while up to 5 ppm of sulphur can be incorporated as Fe/FeS nano inclusions into silicate grains without leaving a significant trace of its presence.}

\authorrunning{K{\"o}hler et al.}
\titlerunning{}

\maketitle


\section{Introduction} 
\label{intro}

The depletion of iron into dust is observed in the diffuse interstellar medium \citep[ISM, see e.g.][]{savage-bohlin-1979,sofia-joseph-1995,jones-2000,jenkins-2009}, but the exact nature of the reservoirs of these elements in dust is unclear. 
Iron can be included into the silicate structure, but observations show that interstellar silicates are mainly Mg-rich.  
A more likely possibility is that iron exists in metallic form incorporated into larger particles as inclusions \citep{constantini-et-al-2005,min-et-al-2007,xiang-et-al-2011,jones-et-al-2013}.
The incorporation of super-paramagnetic inclusions in the form of iron (nano-)particles was for example discussed by \citet[][]{jones-spitzer-1967} to explain the polarisation by dust.
This scenario is also supported by laboratory experiments \citep{davoisne-et-al-2006,djouadi-et-al-2007}.

The abundance and incorporation of sulphur into dust grains is still an open question.
Observations by \citet{ueda-et-al-2005} showed that sulphur is undepleted from the gas phase in the diffuse ISM, while it is highly depleted in cold molecular clouds probably by incorporation into dust grains \citep[see also][]{joseph-et-al-1986, millar-herbst-1990,caselli-et-al-1994}. 
This would imply that sulphur is incorporated into dust grains in the transition from the diffuse ISM to molecular clouds, which is not understood.
Recent studies by U. J. Sofia (private communication) found evidence that sulphur is also depleted in the neutral ISM and that it is therefore an important dust component.
It has been assumed that FeS grains are formed in the protoplanetary disks around young stellar objects \citep{keller-et-al-2002} and also as inclusions in glass with embedded metal and sulphides (GEMS) \citep{bradley-1994}.
\citet{xiang-et-al-2011} and Lee et al. (in prep.) also found that the X-ray dust scattering can best be explained with a dust model containing metallic iron with troilite (FeS).
Furthermore, work on the STARDUST-collected comet 81P/Wild2 dust indicates a mix of glassy silicates and Fe-N-S inclusions in the thermally processed grains \citep{leroux-et-al-2009}.

\begin{SCfigure*}
\includegraphics[width=0.7\textwidth]{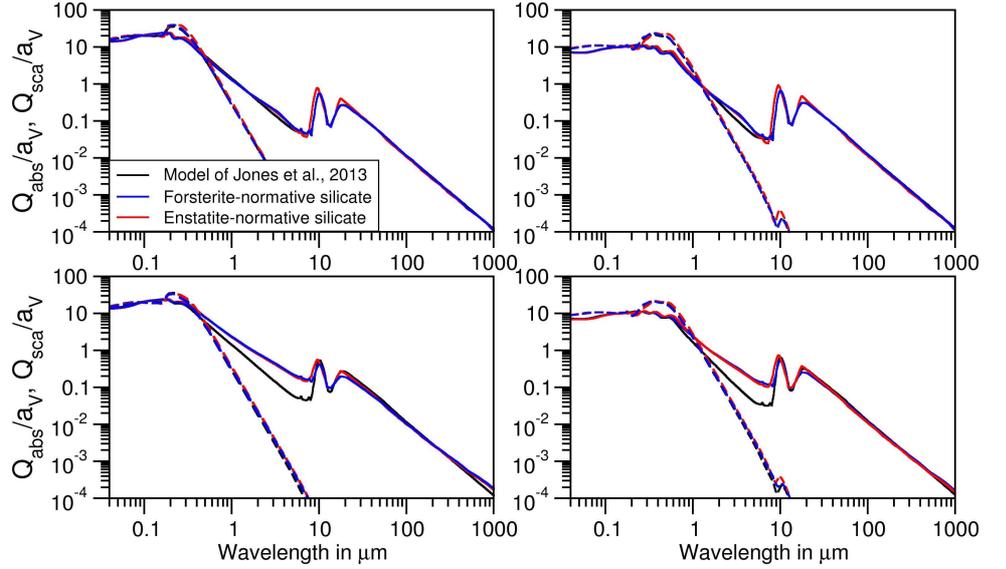}
\caption[]{$Q_{\rm abs}/a$ (solid lines) and $Q_{\rm sca}/a$ (dashed lines) for a=0.06 $\mu$m (left) and a=0.125 $\mu$m (right) for the silicate BGs from the model of \citet{jones-et-al-2013}. Top: compared with the new silicate grains with Fe and FeS inclusions for amorphous silicate with forsterite-normative composition (blue) and enstatite-normative composition (red). Bottom: compared with the same grains with an amorphous-carbon mantle 10 nm thick, instead of 5 nm.}
\label{fig:1}
\end{SCfigure*}

For the diffuse ISM, the dust model of \citet{jones-et-al-2013} accounts for the Fe depletion by including Fe nano-particles in the grains consist of Mg-rich amorphous silicate.
This model considers very small grains (VSGs) and big grains (BGs).
The VSGs consist of aromatic amorphous carbon and BGs consist of core-mantle structures with mantles of aromatic amorphous carbon and cores consisting either of aliphatic amorphous carbon or amorphous silicate of forsterite-normative composition\footnote{Amorphous silicate with the same stoichiometry as forsterite. Equivalently for an enstatite-normative composition.} with 10\% volume Fe metal nano inclusions.
These grains are able to reproduce the observed infrared and UV extinction and spectral energy distribution (SED) of the diffuse ISM.
Small deviations occur only around the 10$-$20 $\mu$m silicate features in the infrared extinction, where the observed features are slightly broader and the extinction around 8 $\mu$m is stronger than in the model.

In the model of \cite{jones-et-al-2013}, the amorphous silicate is assumed to be of forsterite-normative composition.
However, many studies discuss the existence of amorphous silicate with olivine- and pyroxene-normative composition.
In pre-solar grains, \citet{nguyen-et-al-2007} found silicate grains with both compositions.
The dust models reported by \citet{constantini-et-al-2005} and \citet{chiar-tielens-2006} explain observational data by considering both compositions.
The study by \citet{demyk-et-al-2001} showed that evolutionary processes caused by ion irradiation can change an olivine-type amorphous silicate to an amorphous pyroxene-type silicate.
\citet{roskosz-et-al-2011} furthermore showed that for annealing at temperatures below the glass transition temperature, pyroxene-type silicate forms preferentially.
All of these studies suggest that the material composition of the silicate grains is a mixture of amorphous silicates with olivine- and pyroxene-normative compositions.

In this letter, we extend the new dust model of \citet{jones-et-al-2013} by modifying the amorphous silicate BG composition. 
We replace the Fe inclusions with a mixture of Fe and FeS, use amorphous silicate BGs with forsterite- and enstatite-normative composition, and vary the aromatic carbon mantle thickness on the silicate grains.
We calculate the optical properties of the grains with this material composition and compare the results with the original model by \cite{jones-et-al-2013} and observational data.

\section{Material composition and calculations}
\label{sec:3}

We used the Maxwell-Garnett effective-medium theory to derive the optical constants of Fe/FeS inclusions within an amorphous silicate matrix. 
The Maxwell-Garnett mixing rule gives a good approximation of the optical constants of the mixed material.
They consist of a mixture of 70\% Fe \citep{ordal-et-al-1985,ordal-et-al-1988} and of 30\% FeS (troilite) \citep{pollack-et-al-1994} by volume.
In a second step, we used the Maxwell-Garnett rule again to calculate the optical properties for the mixture of 90\% amorphous silicate and 10\% Fe plus FeS mixture. 
We took into account the pair-wise mixing order in the new silicate refractive index data calculation, from most to least refractory materials, as discussed by \citet{jones-2011}.
For the amorphous silicate, we used the optical constants derived by \cite{scott-duley-1996} with forsterite- and enstatite-normative composition. 
We extended the data to longer wavelengths assuming a constant value for $n$ and a decrease of $\lambda^{-1}$ in the slope for $k$. 
These data are consistent with the required Kramers-Kronig relations\footnote{The new silicate data are available on request from the authors.}. 

Subsequently, we used the BHCOAT routine \citep{bohren-huffman-1983} to derive the optical properties of the core-mantle grains. 
The core consists of the mixture of amorphous silicate, either of forsterite- or enstatite-normative composition, and nano-inclusions of the Fe/FeS mixture. 
The mantle consists of aromatic-rich amorphous carbon (band gap energy $E_{\rm g}=0.1$ eV).
The mantle thickness was first assumed to be 5 nm as described by \citet{jones-et-al-2013}. Then in a second calculation we increased the mantle thickness to 10 nm. 
The optical properties for these composite particles were calculated for 30 particle sizes between 0.03 and 5000 $\mu$m to cover the full size distribution.

We used DustEM\footnote{DustEM is a publicly available numerical tool to calculate, for example, the SED, the extinction and the grain temperature distribution for dust in the ISM, available at http://www.ias.u-psud.fr/DUSTEM/.} \citep{compiegne-et-al-2011} to calculate the grain temperatures, extinction, and SED.
First, we considered the model of \citet{jones-et-al-2013} with three types of grains: VSGs of aromatic-rich amorphous carbon up to 20 nm in size, larger grains (BGs) of amorphous carbon with core-mantle structure, where the core consists of aliphatic-rich amorphous carbon and the aromatic-rich amorphous carbon mantle has a thickness of 20 nm, and large grains (BGs) of forsterite-type amorphous silicate with Fe metal nano inclusions with a 5 nm mantle of aromatic-rich amorphous carbon. 
Second, we considered our model with two types of amorphous silicate grains with Fe/FeS inclusions as described above, where we assumed that half of the amorphous silicate grains consist of forsterite-normative composition and the other half of enstatite-normative composition. We furthermore varied the mantle thickness of these silicate grains.
We assumed the same size distribution as described in \citet{jones-et-al-2013} and used the interstellar radiation field (ISRF) described by \citet{mathis-1990} to heat the grains.

In denser regions of the ISM, the individual grains start to coagulate into aggregates.
To understand the influence of the changes on the material composition when coagulating the grains, we carried out calculations using the discrete-dipole approximation (DDA, \citet{purcell-pennypacker-1973, draine-1988, draine-flatau-2010}). 
Aggregates are formed of four BGs and 4000 VSGs. Three BGs consist of amorphous silicate and one BG of amorphous carbon. 
Following the original model of \citet{jones-et-al-2013}, all amorphous silicate BGs have a forsterite-normative compositions and Fe nano-inclusions.
For the new silicates and Fe/FeS inclusions, we replaced the three BGs of amorphous silicate either by two BGs of amorphous silicate with forsterite-normative composition and one BG of amorphous silicate with enstatite-normative composition (model 1) or by one BG of amorphous silicate with forsterite-normative composition and two BGs of amorphous silicate with enstatite-normative composition (model 2). Both types of silicate have 10\% Fe/FeS nano inclusions, as described above.
The method and calculations to derive the optical properties of these aggregates are described in detail in K\"ohler et al. (in prep.). 

\section{Results}
\label{sec:4}

In Fig. \ref{fig:1} we show the absorption and scattering coefficients, $Q_{\rm abs}$ and $Q_{\rm sca}$, divided by the grain radius, $a$, of 0.064 $\mu$m (left) and 0.125 $\mu$m (right) for the amorphous silicate grains of the model of \citet{jones-et-al-2013} and for the new amorphous silicate grains of forsterite and enstatite-normative composition with Fe/FeS inclusions. 
The amorphous-carbon mantle is 5 nm (top) and 10 nm thick (bottom).
The results show only small differences in $Q_{\rm sca}/{a}$.
For a given mantle thickness, we note only small differences in $Q_{\rm abs}/{a}$ between the model of \citet{jones-et-al-2013} and the modified model with forsterite-normative amorphous silicate grains. 
For the amorphous silicate grains of enstatite-normative composition the peak of the silicate features are shifted to shorter wavelengths and the 18 $\mu$m feature is more pronounced. 
The FeS in the Fe/FeS inclusions leads to a slight increase in the absorption in the 3$-$4 $\mu$m region.
Increasing the mantle thickness leads to an increase in $Q_{\rm abs}/{a}$ in the 1$-$10 $\mu$m region compared with the original results of the model of \citet{jones-et-al-2013}, which results in a weakening of the very broad FeS feature.

\begin{figure}[t]
\begin{center}
\includegraphics[width=0.463\textwidth]{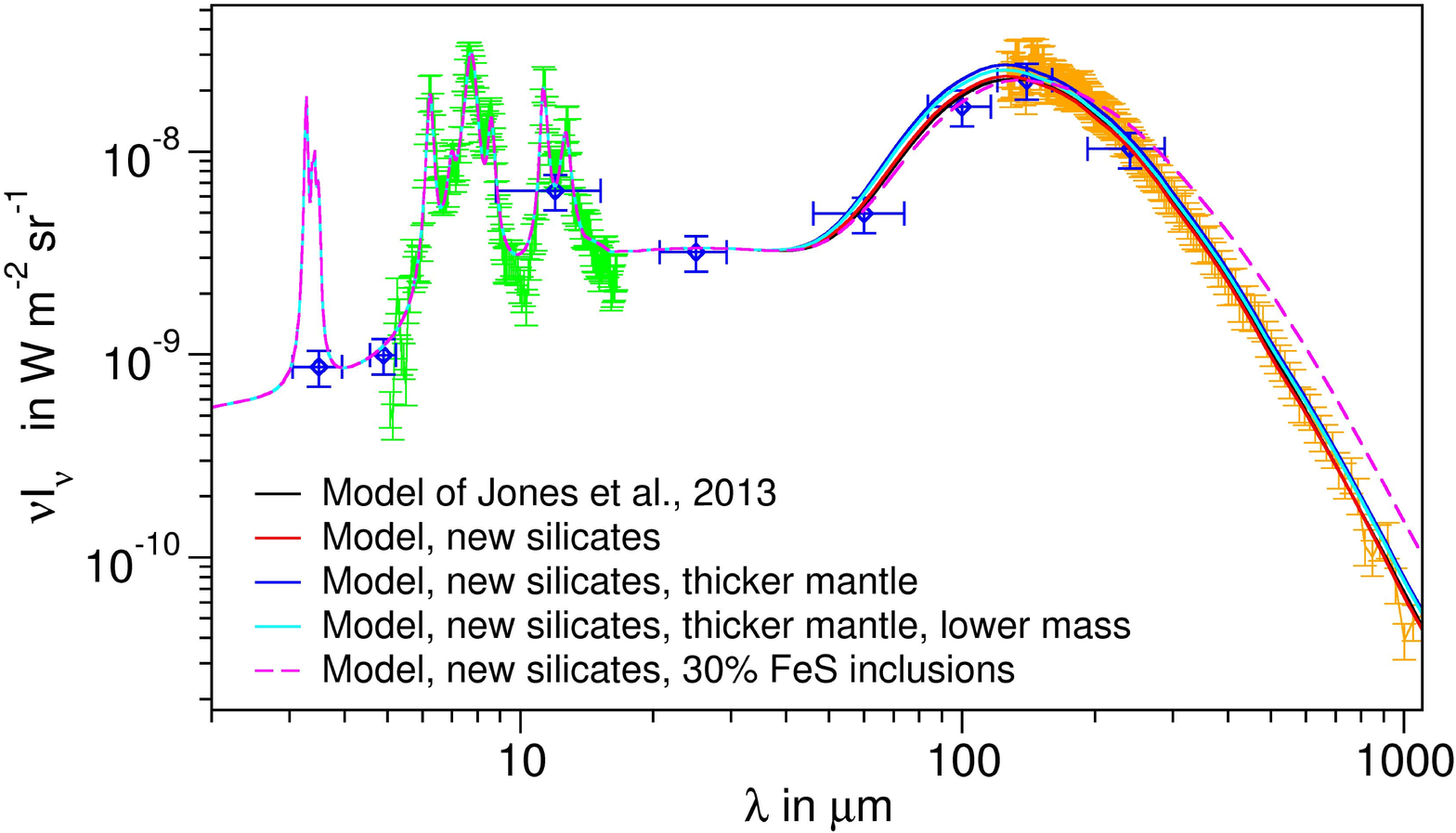}
\caption[]{Dust SED of the diffuse ISM modelled with the original model of \citet{jones-et-al-2013} compared with the SED of the model with new amorphous silicate BGs. Included are observations of the diffuse ISM from ISO ISOCAM/CVF (green curve), COBE FIRAS (orange curve), and COBE DIRBE (blue symbols) \citep[see also][]{compiegne-et-al-2011}.}
\label{fig:2}
\end{center}
\end{figure}

\begin{figure}[t]
\begin{center}
\includegraphics[width=0.45\textwidth]{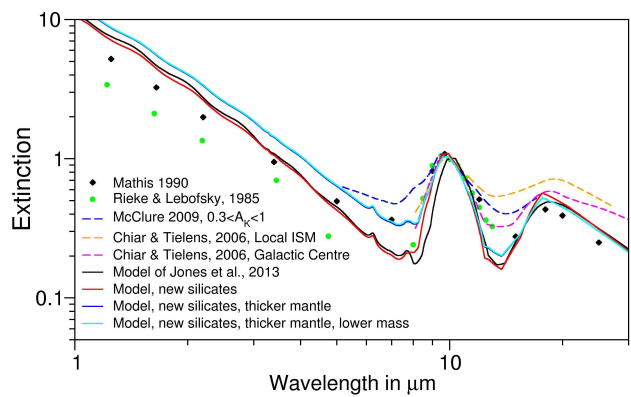}
\caption[]{Extinction of the original model of \citet{jones-et-al-2013} and of the modified model with new amorphous silicate BGs, normalised at 10 $\mu$m, compared with observational data.}
\label{fig:3}
\end{center}
\end{figure}

\begin{figure}[t]
\begin{center}
\includegraphics[width=0.425\textwidth]{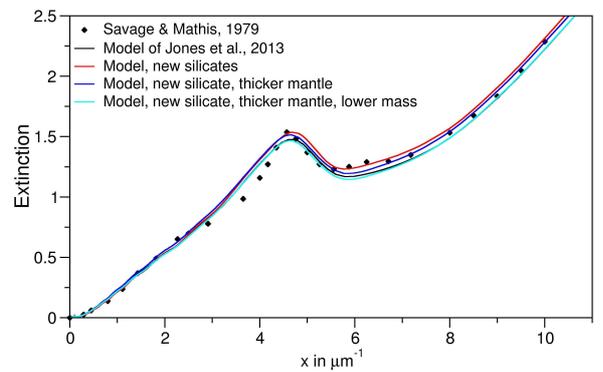}
\caption[]{Extinction in the UV of the original model of \citet{jones-et-al-2013} and of the modified models.}
\label{fig:4}
\end{center}
\end{figure}

The SED and extinction of the diffuse ISM calculated with DustEM are shown in Figs. \ref{fig:2}, \ref{fig:3}, and \ref{fig:4}.
Compared with the original model (black curves), we find only small differences in the SED for the new model (red curves) with the same mantle thickness.
Only for a thick mantle (10 nm) (blue curve) do we find a small increase in emission in the far-IR and a shift of the maximum to shorter wavelengths, indicating an increase in grain temperature.
Reducing the abundance of silicate from $M_{\rm dust}/M_{\rm H}=5.8 \times 10^{-3}$ to $M_{\rm dust}/M_{\rm H}=5.4 \times 10^{-3}$, where each amorphous silicate provides half of the mass and changing the parameter, $a_0$, of the logarithmic normal size distribution as described in \citet{jones-et-al-2013} from 8.0 to 7.0, we again find good agreement with observations (cyan curve).
This requires a silicate element abundance of 31 ppm, which agrees with the assumed abundances in dust \citep[see e.g.][]{compiegne-et-al-2011}. 

In extinction, we find a broadening of the 10 $\mu$m silicate feature caused by the mix of olivine- and pyroxene-type amorphous silicate in the new model (red curve) compared with the original model of \citet{jones-et-al-2013} (black curve) (Fig. \ref{fig:3}), which agrees well with the observational data from \citet{rieke-lebofsky-1986}, \citet{mathis-1990}, and \citet{chiar-tielens-2006}.
The addition of Fe/FeS inclusions into silicates does not change the extinction curve.
The original model and the new model, with a mantle thickness of 5 nm, do not explain the slope of the extinction reported by \citet{mathis-1990} in the 1$-$10 $\mu$m region. 
A mantle of 10 nm leads to an increase in extinction at these wavelengths, which better explains these observations (see blue and cyan curve in Fig. \ref{fig:3}, where data are normalised at 10 $\mu$m). 
However, compared with other observations, for example those by \citet{mcclure-2009}, and at wavelength ranges outside the silicate features we note larger deviations, which cannot be explained with our model. 
We will discuss this extinction problem in detail in a future paper.

Fig. \ref{fig:4} shows that differences in the UV bump and FUV extinction are minimal since they are caused by the VSGs, which we did not change compared to the original model. 
The results of all of the models agree well with the extinction curve of  \citet{savage-mathis-1979}.  

The FeS inclusions in the silicate grains require an S abundance of 2 ppm (cosmic abundance $\sim$16 ppm). 
We find that varying the Fe/FeS mixture up to 1:1 (3 ppm of S) results in only small deviations in the SED and extinction. 
Replacing the Fe/FeS inclusions by pure FeS inclusions (5 ppm of S) shows small deviations in the SED, where the peak shifts to longer wavelengths (decrease in temperature) and the emission at long wavelengths increases. 
Increasing the volume of FeS inclusions to 30\% (13 ppm of S) leads to an increase in the SED at long wavelengths (see pink dashed curve in Fig. \ref{fig:2}).
Increasing the amount of Fe/FeS inclusions to 20\% (3 ppm of S and $\sim$60 ppm of Fe) results in an increase in the extinction in the 1$-$8 $\mu$m region. 
However, such a large increase in the volume fraction of Fe/FeS inclusions is not possible because this is incompatible with the available Fe abundances (cosmic abundance $\sim$32 ppm).
The UV extinction is mainly independent of these differences and we only find small differences in the UV extinction bump.

Pre-solar grains contain SiC with an abundance of $\sim$20 ppm. Including these grains (with an aromatic-rich carbon mantle of 5 nm thickness) does not lead to strong differences in the SED. 
In extinction we find small differences between 2 and 8 $\mu$m and a slight broadening of the silicate features between 10 and 14 $\mu$m.

\begin{figure}[t]
\begin{center}
\includegraphics[width=0.45\textwidth]{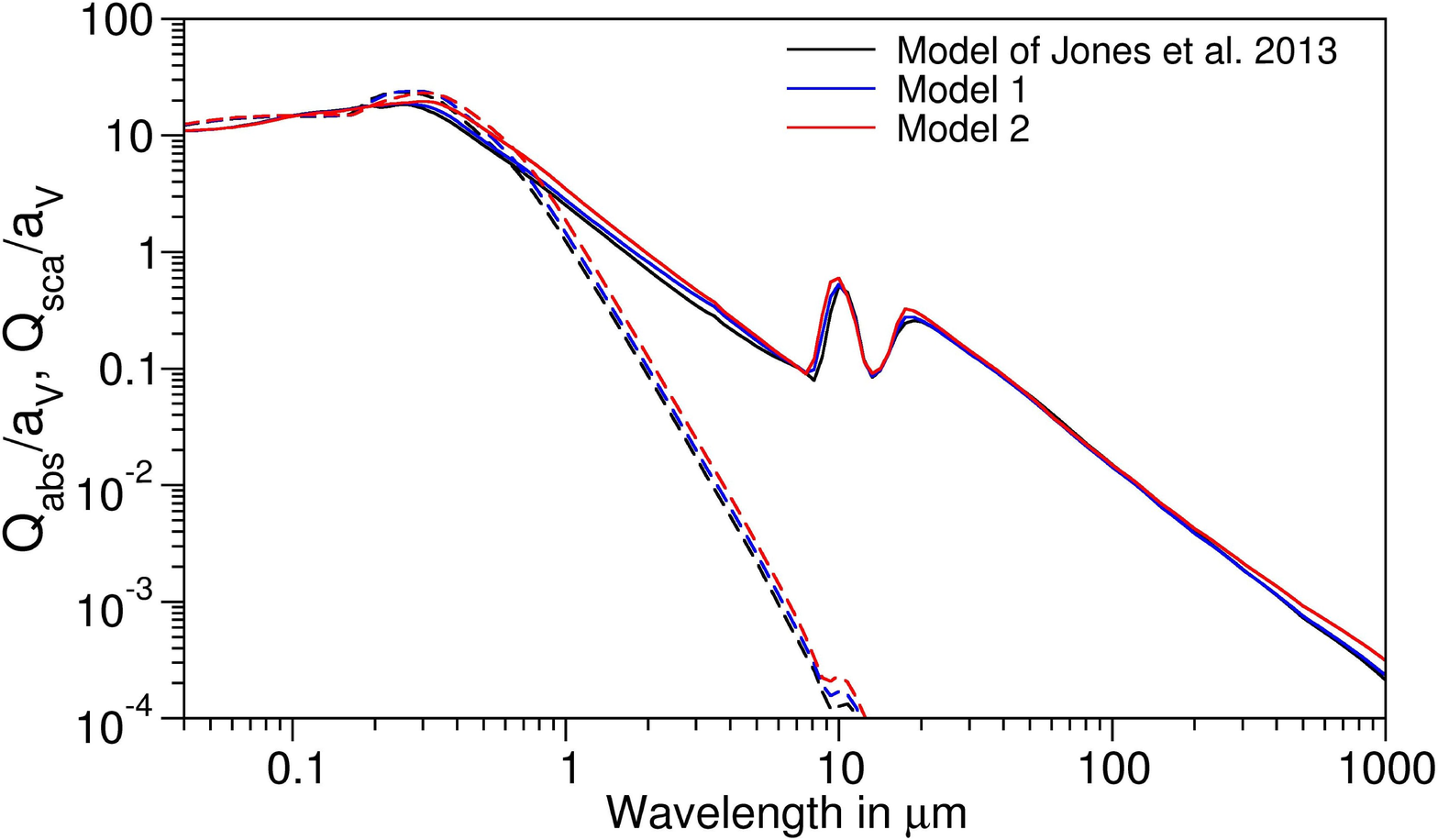}
\caption[]{$Q_{\rm abs}/a_{\rm V}$ (solid lines) and $Q_{\rm sca}/a_{\rm V}$ (dashed lines) for aggregates formed with grains from the original model of \citet{jones-et-al-2013} and for aggregates of models 1 and 2 formed with grains consisting of the new amorphous silicate BGs.}
\label{fig:5}
\end{center}
\end{figure}

In Fig. \ref{fig:5} we present the results for aggregates consisting of the individual grains from the original model by \citet{jones-et-al-2013} (black curve) and for aggregates consisting of the individual grains from the new amorphous silicate composition as described above (blue curve for model 1 and red curve for model 2). 
The optical properties $Q_{\rm abs}/{a_{\rm V}}$ and $Q_{\rm sca}/{a_{\rm V}}$ are shown for one aggregate size, $a_{\rm V}$=0.101 $\mu$m, which is the radius of the volume-equivalent sphere of four single BGs of 0.06 $\mu$m and of 4000 VSGs of 3.5 nm in size. 
The values for $Q_{\rm sca}/{a_{\rm V}}$ are slightly higher for model 2 than model 1.
For model 2 the differences in $Q_{\rm abs}/{a_{\rm V}}$ are larger than for model 1, especially in the 10 and 20 $\mu$m silicate features, which are also slightly broader for model 2, and in the 0.3$-$6 $\mu$m region.
In the V, J, H, and K band, the increase in $Q_{\rm abs}$ for model 2 is a factor of around 1.4.
Whether this effect is grain-size dependent and how it influences the extinction will be discussed in a follow-up paper.
At long wavelengths model 1 does not show strong deviations from the original model by \cite{jones-et-al-2013}, while model 2 shows slightly higher values and a slightly flatter slope.
In summary, for both aggregates and single grains the differences in $Q_{\rm abs}$ and $Q_{\rm sca}$ are small.
In both cases, the mid-IR silicate features are broader, which agrees with observations.

\section{Summary and conclusions}
\label{sec:4}

We have extended the dust model of \citet{jones-et-al-2013} by introducing a mix of amorphous olivine- and pyroxene-type silicate grains with Fe/FeS nano-inclusions and by considering the effects of a thicker carbonaceous mantle.

Adopting a 1:1 mix of amorphous olivine- and pyroxene- type silicates leads to broader silicate features and better agreement with the observed band profiles, in accord with earlier interpretations and the composition of pre-solar silicate grains.

FeS nano-inclusions do not significantly affect the optical properties of amorphous olivine- and pyroxene-type silicate grains/aggregates with Fe metal inclusions. 
We conclude that such composite grains can explain the nature of iron in interstellar dust, that is, as metal inclusions in amorphous silicate, and can provide a reservoir for S (as FeS) in the ISM. 
In particular, we note that FeS inclusions (up to 5 ppm of S) into silicate dust would deplete up to 1/3 of the cosmic S in the ISM in an almost undetectable form.
A larger volume fraction of FeS inclusions results in variations in the SED that appear to be consistent with observations \citep{planck-XI-2013}.

We also showed that increasing the carbon mantle thickness to 10 nm instead of 5 nm enhances the extinction in the 3$-$10 $\mu$m region, which appears to agree better with some observational data. In future studies we plan to investigate this and all of the above effects in more detail.

\begin{acknowledgements} 
The authors would like to thank Julia Lee, U.J. Sofia, Katharina Lodders, and Ed Jenkins for interesting discussions on Fe and S in dust.
This research acknowledges the support of the French Agence National de la Recherche (ANR) through the program ``CIMMES'' (ANR-11-BS56-0029).
\end{acknowledgements}

\bibliographystyle{aa} 
\bibliography{literatur}

\end{document}